\documentclass[fleqn,10pt,twocolumn]{wlscirep}

\usepackage[utf8]{inputenc}
\usepackage[T1]{fontenc}

\usepackage{graphicx} 
\usepackage{dcolumn} 
\usepackage{bm} 
\usepackage{hyperref} 
\usepackage{xr}
\title{Diffusive regimes in a two-dimensional chiral fluid}

\author[1,2]{Francisco Vega Reyes}
\author[1]{Miguel A. L\'opez-Casta\~no}
\author[3,*]{\'Alvaro Rodr\'iguez-Rivas}

\affil[1]{Departamento de F\'{\i}sica, Universidad de Extremadura, 06071 Badajoz, Spain}
\affil[2]{Instituto de Computaci\'on Cient\'{\i}fica Avanzada (ICCAEx), Universidad de Extremadura,
  06071 Badajoz, Spain} \affil[3]{Department of Physical, Chemical and Natural Systems, Pablo de
  Olavide University, 41013, Sevilla, Spain} \affil[*]{arodriguezrivas@upo.es}

\date{\today}

\begin{abstract}
  \textbf{Diffusion is a fundamental aspect of transport processes in biological systems, and thus,
    in the development of life itself. And yet, the diffusive dynamics of active fluids with
    directed rotation, known as chiral fluids, has not been analyzed in detail so far. Here, we
    describe the diffusive regimes of a two-dimensional chiral fluid, composed in this case of a set
    of identical disk-shaped rotors. We found strong experimental evidence of odd diffusion. This
    odd diffusion emerges in the form of a two-dimensional tensor with an antisymmetric part. In
    particular, we show that chiral diffusion is complex, featuring transitions between super,
    quasi-normal, and sub diffusion, and very slowly aging. Moreover, we show that the diffusion
    tensor elements, including off-diagonal elements; i.e., odd diffusion coefficient, change sign
    according to flow vorticity. Therefore, the chiral fluid has a self regulated diffusion,
    controlled by its vorticity. }
\end{abstract}

\begin{document}
\flushbottom
\maketitle

\thispagestyle{empty}
\noindent

\section*{Introduction}

Chiral fluids display a rather complex dynamics, featuring a variety of peculiar properties that
cannot be found in traditional fluids. Albeit these properties play actually a major role in
biological processes \cite{MJC16}, where flow chirality is ubiquitous \cite{VZ12}, research work on
this line is currently in progress \cite{WHB11,PWL15,BC16,WL00,ENCGZ22,ZYSOS22}.  In fact, the
hydrodynamics of chiral fluids is characterized by an antisymmetric component in the stress tensor,
and by a whole new set of transport coefficients. These new coefficients, allusively dubbed as odd
(due to their chiral origin \cite{A98}) include new contributions (for instance, to viscosity
\cite{A98,Banerjee2017,HFSVPV21}). Diffusion is illustrative in that, whereas it has been
extensively studied in a variety of materials since long ago \cite{KCSSM21,CP20,GCG16,KA94}, the
development of diffusion theory for chiral fluids is relatively recent \cite{HEM21}. For instance,
no experimental evidence of odd diffusion in a chiral fluid of active rotors has been provided so
far (although there are experimental evidences in colloids under magnetic fields)
\cite{TH08,ZSS20,RAV21,BUGPSMSBI22}. Furthermore, in the present work, we provide strong
experimental evidence of a rich and complex diffusive behaviour.

Parity and time reversal symmetry break imply the emergence of particle velocity cross correlations
\cite{LMMRV21}, which necessarily yields, as we will see, an asymmetric diffusion tensor

\begin{equation}
  \label{eq:diffusion_tensor}
  \mathcal{D} = \begin{bmatrix}
    D &  -D^\mathrm{odd}\\
    D^\mathrm{odd} & D   
\end{bmatrix},
\end{equation} where each of the two coefficients appearing in \eqref{eq:diffusion_tensor},
including those corresponding to an antisymmetric component, $D^\mathrm{odd}$, can be calculated
according to the Green-Kubo \cite{Green1954,Kubo1957,HEM21,HKEM20} relations

\begin{align}
  \label{eq:GK_diff}
  D &= \frac{1}{2}\int_{0}^{\infty}\mathrm{d}t\,\langle v_i(t)v_j(0)\delta_{ij}\rangle, \\
  D^\mathrm{odd}&= -\frac{1}{2}\int_{0}^{\infty}\mathrm{d}t\,\langle v_i(t)v_j(0)\epsilon_{ij}\rangle,
  \label{eq:GK_odd_diff}
\end{align} with $\delta_{ij}, \epsilon_{ij}$ being the Kronecker delta and the 2D Levi-Civita
symbol, respectively. In~\eqref{eq:GK_diff}, diagonal elements ($D$) represent the Brownian diffusion
coefficient, whereas off-diagonal elements ($D^\mathrm{odd}$) stand for what we denoted as odd
  diffusion coefficient, in close analogy with odd viscosity \cite{A98}. 

  A recent theoretical model has predicted a diffusion tensor of the form in
  Equation~\eqref{eq:diffusion_tensor} as well \cite{HEM21}. However, in this model, the sign of
  $D^\mathrm{odd}$ is enslaved to the sign of particle activity. By contrast, we show now that
  chiral diffusion is governed, at experimental level, by chiral flow vorticity (thus, by a fluid
  property, not a particle property). Furthermore, we describe a peculiar diffusive behavior. Beyond
  the question of the existence of the odd diffusion coefficient $D^\mathrm{odd}$, we also found
  that: i) chiral diffusion is always very slowly aging, and ii) the chiral fluid is in general
  super-diffusive but can decay to normal diffusion or sub-diffusive behavior, with these
  transitions being controlled by flow vorticity as well.


\section*{Results}
\label{sec:results}

\subsection*{Description of the system}
\label{sec:setup}
We performed experiments with a prototypical chiral fluid system. Here, the fluid consists of a set
of identical disks-shaped particles. The disk diameter is $\sigma=7.25~\mathrm{cm}$, mass is
$m_p=7.1~\mathrm{g}$, and $I\simeq (1/8)~m_p\,\sigma^2=46~\mathrm{g\,cm^2}$ is the moment of
inertia. Each particle features a set of 14 evenly spaced, equally tilted blades. This geometrical
configuration yields an intrinsic rotational chirality to the particles, which results in stationary
chiral flows \cite{LMMRV21}. The disks translations are driven by turbulent air wakes produced by a
homogeneous air upflow past the particles (for a more complete description of the driving mechanism,
see \cite{LMMRV21}). This upflow also generates a continuous disk spin, as it goes through the
blades. Due to their blade tilt angle, our particles spin clockwise (i.e., negative spin, according
to mathematical convention). The disks are confined within a horizontal circular region of diameter
$L=72.5~\mathrm{cm}$. Each experiment is recorded for $100~\mathrm{s}$ with high speed video
camera. From experiments movies we obtain, by means of a particle tracking algorithm \cite{CG96},
the data presented in this work. See the Methods section for more details on the experimental
methods.

Our analysis will be based on the spatial average of the fluid vorticity, $\overline\omega$. Here,
we define flow vorticity as $\omega=(1/2)\epsilon_{ij}\partial_i u_j$. (As we will later see,
$\overline\omega$ controls the diffusive behavior.) It is also convenient to define the packing
fraction as $\phi\equiv N(\sigma/L)^2$, where $N$ is the number of rotors present in the system (in
our sets of measurements, $20<N<75$).


\begin{center}
  \begin{figure}[ht!]
	\includegraphics[width=0.99\columnwidth]{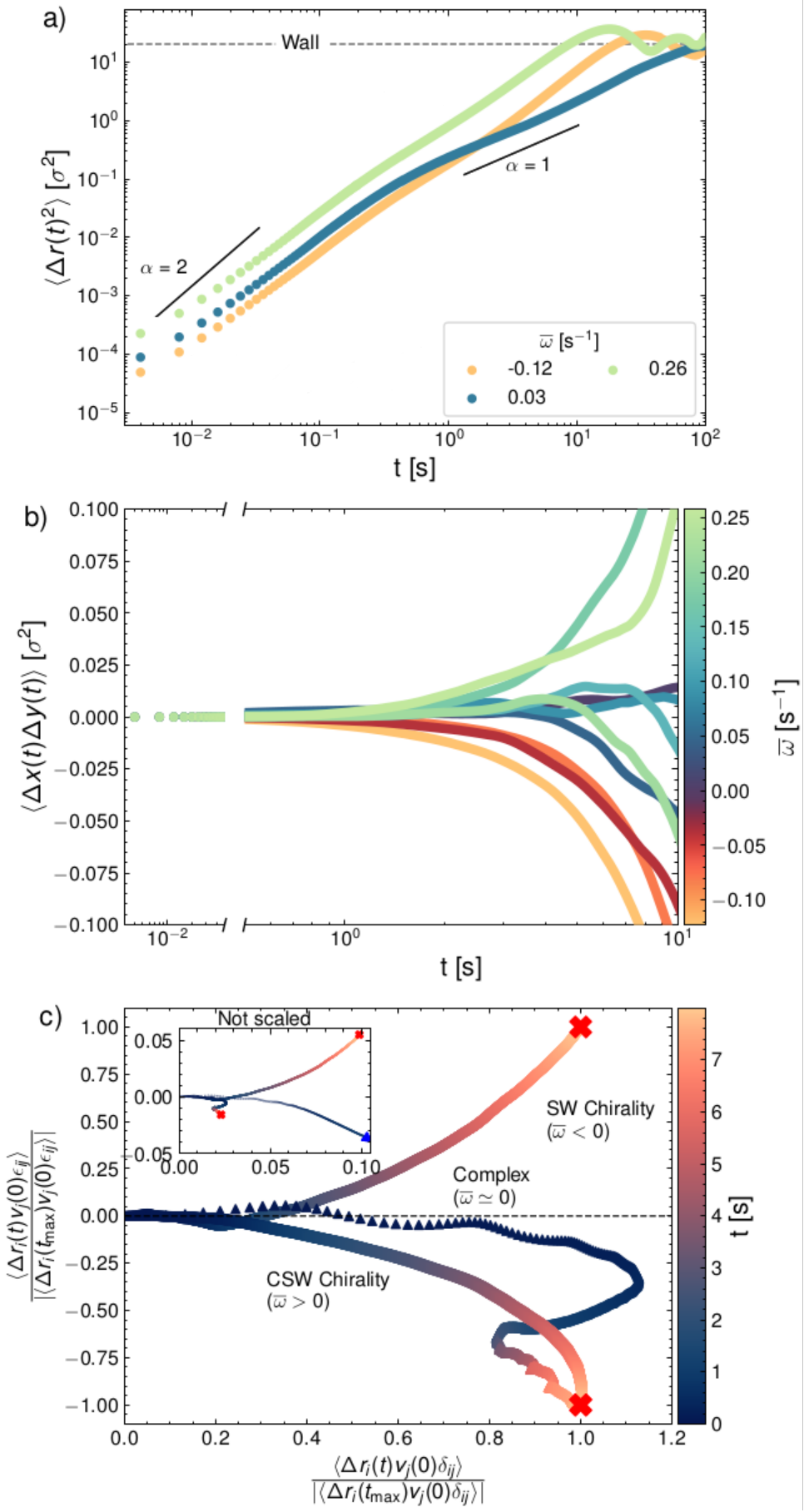}
	\caption{\textbf{Statistical properties of particle displacements.} (a) Mean squared
          displacement $\Delta r(t)^2$ at three different global vorticities $\overline\omega$.  The
          saturation value due to confinement is indicated with a dashed line.  (b) Mean
          cross-displacements $\langle\Delta x(t)\Delta y(t)\rangle$ for several values of
          $\overline\omega$. (c) Position-velocity cross correlations for
          $\overline\omega=-0.12~\mathrm{s}^{-1}$, $\overline\omega=0.03~\mathrm{s}^{-1}$,
          $\overline\omega=0.26~\mathrm{s}^{-1}$ (indicated as $\overline\omega<0$,
          $\overline\omega=0$, $\overline\omega>0$ respectively). In the main panel of (c), correlations
          are scaled with the corresponding value at $t = t_{\mathrm{max}}=8~\mathrm{s}$ (the last
          represented value in this case), while in inset shows the unscaled version. Packing
          fraction for (a)-(c): $\phi=0.45$.}
	\label{fig:displacements}
      \end{figure}
\end{center}

\begin{center}
	\begin{figure}
          \centering \includegraphics[width=0.75\columnwidth]{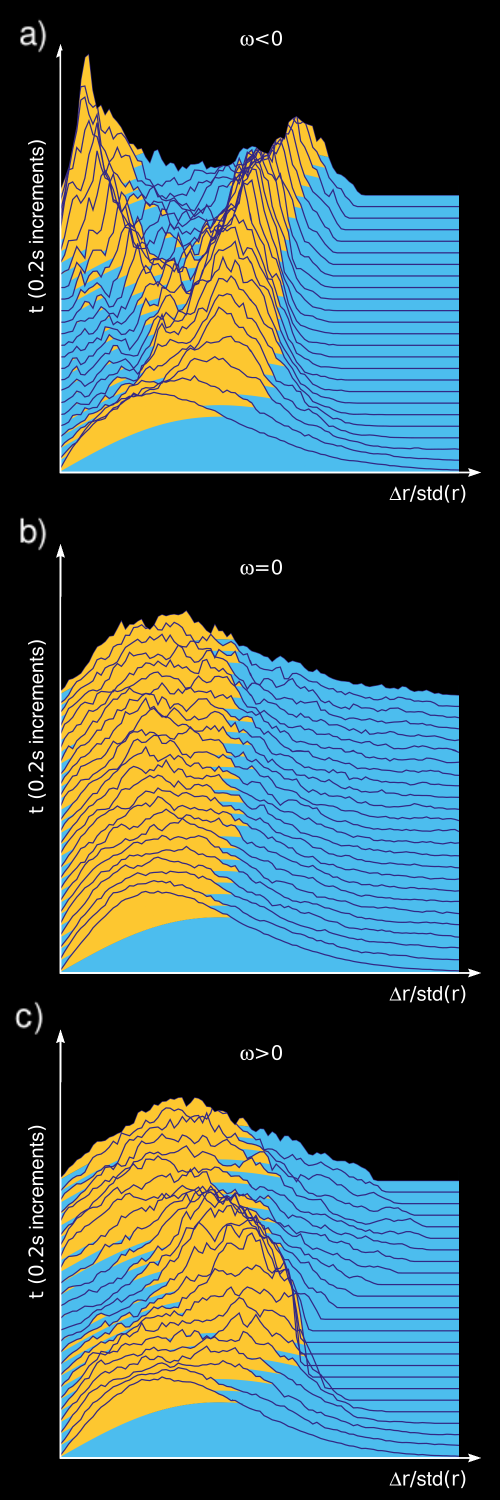}
          \caption{\textbf{Time evolution of the distribution function of reduced mean square
              displacements, $f(\Delta r/\mathrm{std}(\Delta r))$}. We
            display results for (a) $\overline\omega=-0.15~\mathrm{s}^{-1}$, (b)
            $\overline\omega=0.08~\mathrm{s}^{-1}$ and (c) $\overline\omega=0.23~\mathrm{s}^{-1}$
            (indicated as $\overline\omega<0$, $\overline\omega=0$, $\overline\omega>0$
            respectively), with packing fraction $\phi=0.45$ . Being $\mathrm{std}(\Delta r)$
            the standard deviation of $\Delta\mathbf{r}$ . In orange, the areas where
            $f(\Delta r/\mathrm{std}(\Delta r))>f_\mathrm{G}$ and blue where
            $f(\Delta r/\mathrm{std}(\Delta r))<f_\mathrm{G}$ (here, $f_\mathrm{G}$ is the normal
            Gaussian distribution).}
          \label{fig:distributions}
              \end{figure}
\end{center}

\subsection*{Mean squared and cross displacements}
\label{subsec:displacements}

In Figure 1 we analyze the properties of the averages of particle displacements. In particular,
Figure~\ref{fig:displacements}~(a) shows the ensemble mean squared displacement
$\langle\Delta r(t)^2\rangle\equiv\langle\Delta x(t)^2 + \Delta y(t)^2\rangle$, where
$\Delta x(t)^2\equiv (1/\mathcal{N}(t))\sum_{\{t_0\}}[x(t+t_0)-x(t_0)]^2$, for positive, negative
and zero global vorticity. (Since we deal with steady states only, displacements can be obtained,
for each lag time $t$, from an average over the $\mathcal{N}(t)$ available initial states $t_0$.) In
logarithmic scale, $\langle\Delta r(t)^2\rangle$ grows until confinement effects enter into action
(particles should display diffusive arrest at, approximately, half the system size \cite{B07}). As
we see, the slope (in logarithmic scale) of $\langle\Delta r(t)^2\rangle$ remains slightly varying
at all times; i. e., never behaves quite linearly enough, contrary to the linear behavior in normal
diffusion \cite{MJCB14}. In this figure we represent two slopes ($\alpha=1$, $\alpha=2$), making it
clear that the system is superdiffusive ($2>\alpha>1$). Furthermore, in most cases, the slope never
reaches the typical value for normal diffusion, that is, $\alpha=1$.

Figure~\ref{fig:displacements}~(b) displays the corresponding time evolution of the ensemble mean
cross displacement $\langle\Delta x(t) \Delta y(t)\rangle$. With
$\Delta x(t) \Delta y(t) = (1/\mathcal{N}(t))\sum_{\{t_0\}}[x(t+t_0)-x(t_0)][y(t+t_0)-y(t_0)]$. This
ensemble average would in general be zero in a system without chirality, and for this reason is
usually not measured \cite{MJCB14}. Here, however, $\langle\Delta x(t) \Delta y(t)\rangle\neq 0$ in
general. Nevertheless, we observe that $\langle\Delta x(t) \Delta y(t)\rangle$ remains null during a
significatively long time interval. Furthermore, for experiments with $\overline\omega\simeq 0$,
$\langle\Delta x(t) \Delta y(t)\rangle$ is nearly null at all times. Otherwise, the mean cross
displacement is monotonically increasing/decreasing for $\overline\omega>0$, $\overline\omega<0$
respectively. This property has been measured during a time shorter than one fluid complete
revolution but, in any case, a wide range of values is observed within this interval. This provides
rich information on the behavior in the relevant parameter space. The dynamics of particle
displacements is shown for more clarity also in the Supplementary Movie 1.

Finally, in panel (c) of Figure~\ref{fig:displacements}, we see a spiral-like behavior of
$\langle \Delta r_i(t) v_j(0)\epsilon_{ij}\rangle$ vs.
$\langle\Delta r_i(t)\Delta v_j(0)\delta_{ij}\rangle$, for $\overline\omega\neq0$. The shape of
these spirals indicate that, in effect, we observe strong odd diffusion in general, but with
important departures from results reported previously in theoretical models
\cite{TH08,NEGL16,HEM21}. In effect, we notice that, to the best of our knowledge, a new feature
appears here, in the form of a kink with reverse curvature that develops at early times, near the
transitional region $\overline\omega\simeq 0$. As $\overline\omega$ decreases, this kink with
reverse curvature becomes bigger, which results in a combination of two opposite curvature
spirals. The kink eventually occupies the complete spiral which, by means of this continuous
mechanism, reverses its curvature sign (we point out here that the typical measurement error is 100
times smaller than the size of the kink and thus this feature is captured here with a high level of
accuracy). As we said, this observation suggests that the change of sign of the odd diffusion
coefficients occurs gradually. Moreover, it appears to be mediated by a (to our knowledge) not
previously reported behavior where the odd coefficient alternatively changes its sign as time
evolves (i.e. in some chiral stationary flows, the sign of the odd diffusion is not clearly
defined). Supplementary Figure 1 shows he transitions of the spiral curvatures and evolution of
their kinks in more detail.


In order to provide also a description of the microscopic structure of diffusion, we represent in
Figure~\ref{fig:distributions} the distribution functions of the root of the mean squared
displacements, $\Delta r(t)$, reduced with the standard deviation $\mathrm{std}(\Delta r (t))$. We
plot the results for a set of lag times, so that we can track eventual deviations off the Gaussian
distribution (i.e., deviations from normal diffusion \cite{MJCB14}). We can detect strong deviations
from normal diffusion for the $\overline\omega<0$ (Figure~\ref{fig:distributions} a) and
$\overline\omega>0$ (Figure~\ref{fig:distributions} c) cases. Moreover, these deviations slowly vary
as the system ages. Also, deviations from Gaussian behavior bifurcate for $\overline\omega<0$ at
longer lag times, see panel (a) (and also Supplementary Movie 1). On the contrary,
$f(\Delta r/\mathrm{std}(\Delta r))$ deviations from the Gaussian preserve the same structure at all
times for $\overline\omega\simeq 0$ and they occur in the range of short displacements only, thus
indicating that diffusion is less complex in this case. For more detail, see also Supplementary
Figure 2 for a three-dimensional representation of the excess kurtosis of
$f(\Delta r/\mathrm{std}(\Delta r))$. We found that that $f(\Delta r/\mathrm{std}(\Delta r))$ is
always platykurtic; i.e., the displacements distributions inherently have thinner
tails. Additionally, we have observed that the distribution function of cross displacements is
clearly asymmetric when a strong chiral flow is present (i.e., when $|\overline\omega|$ is not
small), which seems to indicate also that the chiral flow is the origin of odd diffusion in our
system. See Supplementary Figure 3 for details on the behavior of $f(\Delta x(t)\Delta y(t))$.

\begin{figure*}[ht!]
  \includegraphics[width=0.99\textwidth]{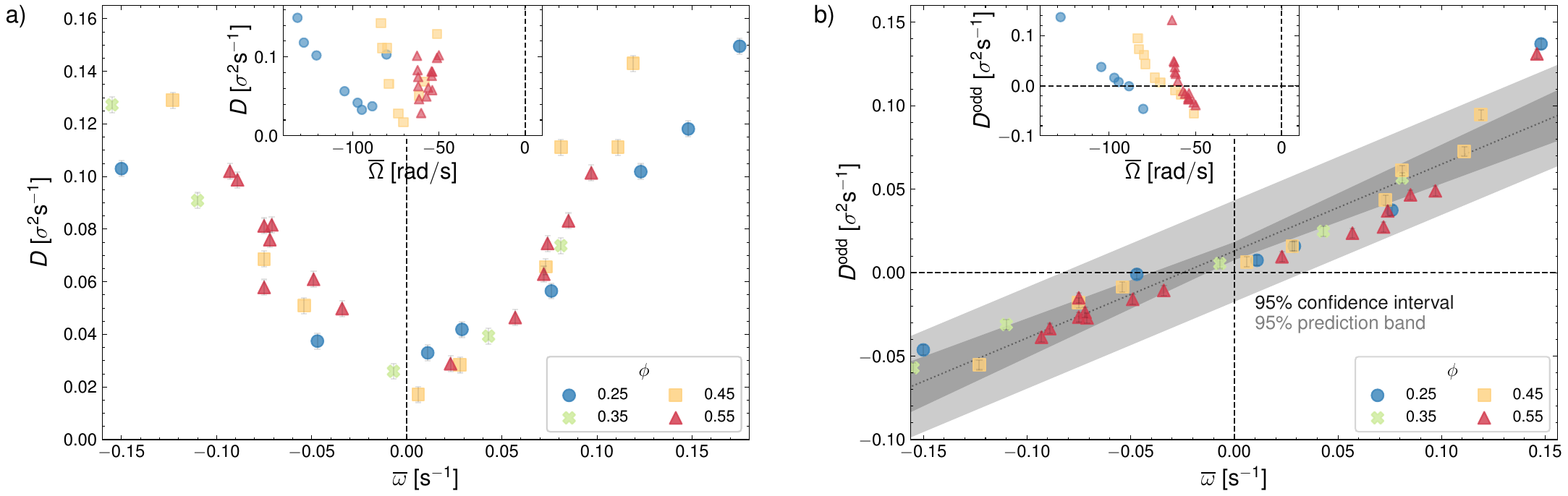}
  \caption{\textbf{Diffusion coefficients $D$, $D^\mathrm{odd}$ vs. global mean vorticity
      $\overline\omega$}. Results are represented for several experiments at different densities
    (each denoted with a different symbol and color, see figure legend) (a) Diffusion coefficient
    $D$ (diagonal elements of diffusion tensor $\mathcal{D}$, in
    Equation~\eqref{eq:diffusion_tensor}). (b) Odd diffusion $D^\mathrm{odd}$ coefficient
    (off-diagonal elements of $\mathcal{D}$). Confidence intervals of a linear regression from the
    experimental data for $D^\mathrm{odd}$ appear shaded in grey. Insets display diffusion
    coefficients vs. particle activity (here, average particle spin $\overline\Omega$). Notice that
    in the insets $D^\mathrm{odd}$ changes sign at different points, contrary to the behavior with
    respect to the true control parameter $\overline\omega$ (main panels).}
  \label{fig:diff_tensor}
\end{figure*}


\subsection*{Diffusion coefficients}
\label{subsec:diffcoeff}
The results for the diffusion coefficients $D,~D^\mathrm{odd}$, as obtained from
Equations~\eqref{eq:GK_diff},\ \eqref{eq:GK_odd_diff}, are shown in Figure~\ref{fig:diff_tensor}. It
is to be noticed that odd diffusion (right panel) is quite significant in most of the parameter
space, since $D^\mathrm{odd}$ is close to the order of magnitude of $D$ (left panel), except in the
region of $\overline \omega \simeq 0$, where $D^\mathrm{odd}$ systematically falls to
zero. Furthermore, Figure~\ref{fig:diff_tensor} clearly reveals that $\overline\omega=0$ signals in
all cases the points where $D^\mathrm{odd}=0$ and the absolute minimum of $D$ (and hence, the trend
inversion of $D$). Moreover, all the curves for different densities collapse into one when
represented against global vorticity, as an additional strong experimental evidence that, in effect,
global vorticity $\overline\omega$ is the true control parameter for the diffusive regimes in a
two-dimensional chiral fluid. This means that diffusion in the two-dimensional fluid is controlled
by the global vorticity of the chiral flow; i.e. a fluid property, and not a microscopic property
(such as particle activity/chirality \cite{TH08,HEM21}). This can be best seen in the insets of
Figure~\ref{fig:diff_tensor}, where the coefficients are also represented as a function of particle
activity (here, the ensemble average spin, denoted as $\overline\Omega$). Furthermore, although the
system is locally heterogeneous, we have observed that the sign of $\overline\omega$ is the same
throughout the system\cite{LMMRV21}; i.e., the sign of chiral flow vorticity is a global property of
the system (except for $\overline \omega \simeq 0$, where the behavior becomes more complex). This
allows us for spatially averaging the diffusion tensor components, which aids to detect the relevant
regimes in this kind of complex dynamics.


\begin{center}
  \begin{figure}[ht!]
  \includegraphics[width=0.99\columnwidth]{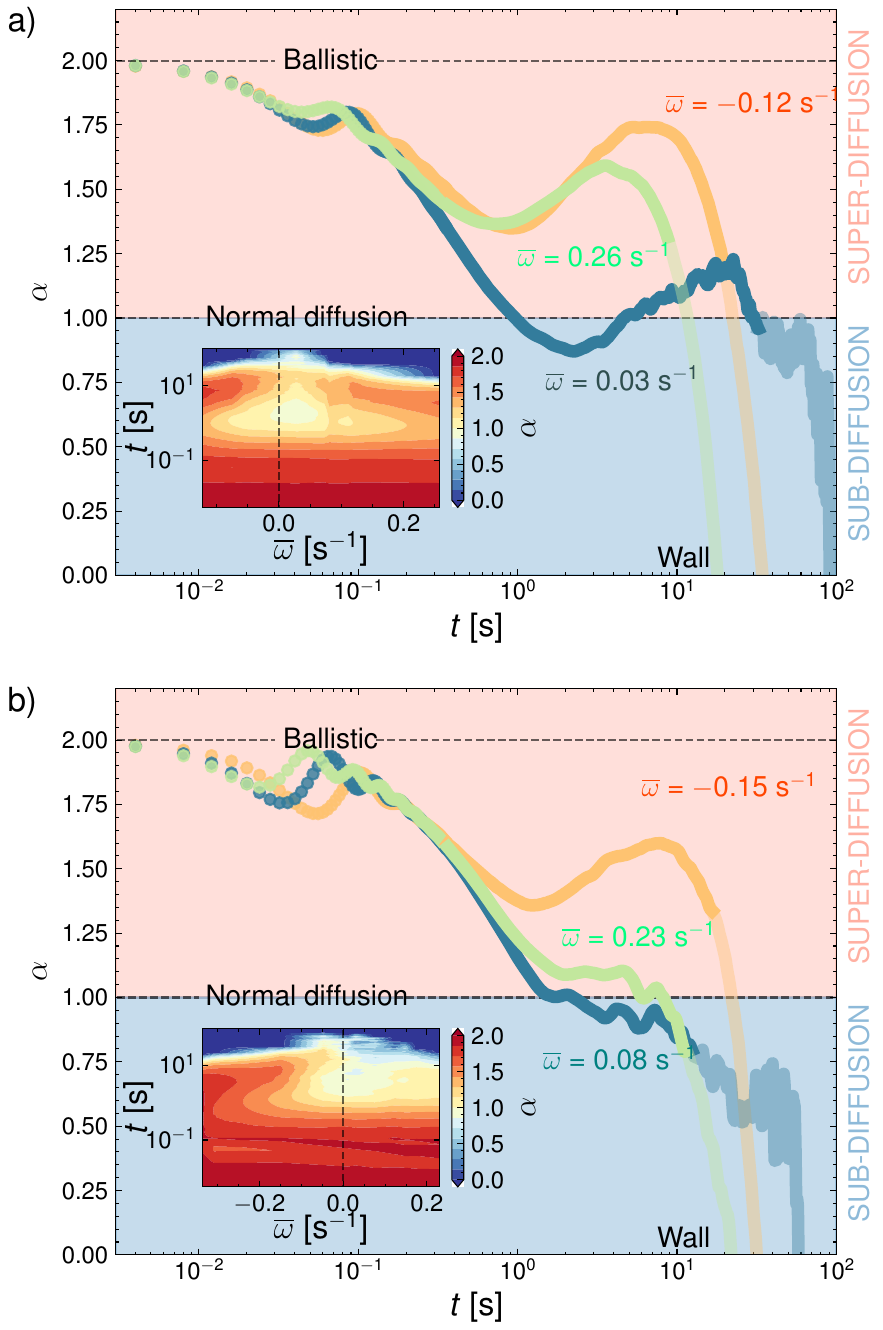}
  \caption{\textbf{Evolution of the diffusive exponent $\alpha$}. We show the results for three experiments with 
    $\overline\omega<0, \overline\omega=0, \overline\omega>0$ respectively 
    (each represented in a different color of curve), at two different packing 
    fractions: (a) $\phi=0.45$ and (b) $\phi=0.25$. 
    Here, $t$ represents lag time. The diffusive exponent has been calculated by performing a rolling fit of the
    mean squared displacement to a $t^{\alpha}$ slope. The ballistic regime and normal diffusion
    regimes are signaled with horizontal lines at $\alpha=2$ and $\alpha=1$,
    respectively. Translucent lines indicate that the system boundary has been reached. Heat-maps in
    the insets represent $\alpha$ in the $\overline\omega\text{--}t$ plane (notice the dip at large
    $t$, for the case $\overline{\omega} \simeq 0$).}
  \label{fig:alpha}
\end{figure}
\end{center}


Results from Figures~\ref{fig:displacements}~(a) and~\ref{fig:distributions} are actually revealing
that mean square displacements for chiral fluids do not show in general (except, and only
approximately, for the cases with $\overline\omega\simeq 0$) the typically exponential behavior of
both normal and anomalous diffusion \cite{MJCB14}. Additionally, it is also apparent that the slope
of the corresponding curve $\langle\Delta r(t)^2\rangle$ vs. $t$ in log scale is time varying, as we
said. This is in fact in close analogy with the behavior of jammed granular systems
\cite{AD06,LGRAYV21} (indicates a varying exponential diffusion coefficient, as it happens in
granular jamming \cite{AD06}). 

In order to analyze this feature in more detail, we plot in Figure~\ref{fig:alpha} the time
evolution of the diffusive exponent ($\alpha$) of $D$, defined \cite{MJCB14,LGRAYV21} as
$\langle\Delta r(t)^2\rangle=(4D)t^\alpha$.

All $3$ cases $\overline{\omega}<0$, $\overline{\omega}\simeq0$, $\overline{\omega}>0$ undergo a
slow decay off the initial ballistic regime ($\alpha=2$) at short times (slow refers here at time
intervals much longer in the diffusion curves than the ballistic regime, this is shorter than $0.01$
s). Moreover, at high densities ($\phi=0.45$, Figure~\ref{fig:alpha} a), the case
$\overline\omega\simeq 0$ has a diffusive exponent $\alpha$ that slowly decays to nearly normal
diffusion ($\alpha=1$). On the contrary, the other two cases ($\overline\omega\neq 0$) remain
superdiffusive, with large oscillations around a mean value $\alpha \sim 1.5$ at high
densities. Consequently, the inset of Figure~\ref{fig:alpha}~(a) reveals strong superdiffusion in a
large fraction of the parameter space, in this case for a system with packing fraction
$\phi=0.45$. However, at low densities ($\phi=0.25$, Figure~\ref{fig:alpha}~b) the phase behavior of
diffusion seems to be qualitatively different. Now the system displays normal or weakly subdiffusive
behavior for $\overline\omega\simeq 0$ and $\overline\omega>0$ cases, whereas the
$\overline\omega<0$ case remains strongly superdiffusive (it would require to go to much larger
densities in order to detect jamming or glassy behaviour). In all cases, $\alpha$ sharply falls to
$\alpha=0$ at very long times, due to the effect of the wall confinement \cite{MJCB14}.

\section*{Discussion}

In summary, we have provided an exhaustive description of the experimental behavior of chiral
diffusion in a two-dimensional fluid of active particles. It is worth to remark that an odd
diffusion coefficient $D^\mathrm{odd}$ is indeed experimentally measurable in chiral flows. However,
and although $D^\mathrm{odd}$ is frequently as large as the usual diffusion coefficient $D$, our
experiments reveal that odd diffusion can also be absent for active particles (i.e., at non-zero
particle activity, see inset in Figure~\ref{fig:diff_tensor}~b), contrary to what has been recently
predicted \cite{HEM21}. Thus, more theoretical analysis should be performed on this important
question.

Moreover, we find as well the following chiral diffusion regimes: I) diffusion
coefficient $D$ is superdiffusive, accompanied with large odd diffusion coefficient $D^\mathrm{odd}$
(predominant regime); II) diffusion coefficient is weakly subdiffusive, with large odd diffusion
coefficient $D^\mathrm{odd}$; III) the chiral fluid displays quasi-normal diffusion,
with vanishing odd diffusion $D^\mathrm{odd}$. For all three diffusive regimes, the diffusion
coefficient $D$ decays very slowly from ballistic to diffusive regime, which is prone to the
emergence of memory effects \cite{LVPS17}. Furthermore, in the few cases where the system becomes
subdiffusive (see Figure~\ref{fig:alpha} b), the logarithmic slope of displacements changes
continuously (Figure~\ref{fig:displacements} a) and is not accompanied by a plateau, as it happens
in molecular \cite{RRR19} and granular glasses \cite{LGRAYV21} (where subdiffusive particles are
caged by the neighbors and only at long times, diffusing from a cage to the next); i.e. subdiffusion
in chiral fluids appears to be rather peculiar with respect to their analogs. It is important to
highlight that a very recent study of the structure factor by means of simulation shows
locally-jammed states and multi-scale hyperuniformities for unconfined active spinners\cite{RJMK22},
which may be related to the weakly subdiffusive behavior that we observe here.

Additionally, Figure~\ref{fig:displacements}~(b) shows that the time evolution of the mean cross
displacement remains null for early to mid development stages, and
emerging only (when it does) as the trajectories age enough, which results in a peculiar structure
of the Brownian movement of the particles in the chiral fluid, as illustrated in
Figure~\ref{fig:displacements} (and, more directly, in Supplementary Figure $4$).

Most importantly, we have discovered that the control parameter of chiral diffusion is the flow
vorticity $\overline\omega$. This is supported by the strong and first (to the best of our
knowledge) experimental evidence of the universal curves of the diffusion coefficients
vs. $\overline\omega$ in Figure~\ref{fig:diff_tensor}. It actually makes sense that diffusion does
not depend on (chiral) particle activity alone, since one could expect that the underlying chiral
fluid flow has a strong impact on diffusion processes. In fact, this agrees very well with the
recent experimental evidence that flow vorticity is not controlled by particle activity either
\cite{LMMRV21}. Furthermore, the transitions between the aforementioned three diffusive regimes were
still detected in spite of nearly constant reduced activity (see Supplementary Figure $5$, where
insets of Figure~\ref{fig:diff_tensor} are represented with more detail), which renders impossible
that reduced particle activity can control chiral and hence odd diffusion.

This fundamental feature of odd diffusion in our system, and most of the other ones
described in the present work, had not been reported previously and in fact could not be trivially
expected from a theoretical point of view. Our results are actually more reminiscent of more
realistic biological systems with non-lattice (directionally) correlated random walks
\cite{CPB08}. It remains for future work the development of a theoretical framework that can account
for the mechanisms causing these peculiar features of chiral diffusion in two dimensions in
experimental systems. We are currently making progress in this sense.

In a broader context, the behaviors described here altogether draw a landscape of diffusion for
two-dimensional chiral fluids that is very complex and peculiar, and probably unique in nature. We
think this can potentially help to describe important dynamical processes in biologic systems.

\section*{Methods}

In our experiments we use $N$ identical 3D-printed disks (.stl models are available upon request),
made of polylactic acid (PLA). Each disk has 14 equally-spaced identical oblique blades that
generate a clock-wise spinning as upflow passes through the disk. Turbulent vortexes from the upflow
past the particles produces a stochastic translational motion as well. The rotors are located on top
of a perforated steel sheet ($3~\mathrm{mm}$ diameter holes in a hexagonal pattern, with a
$3~\mathrm{mm}$ spacing). This metallic grid is mounted on a box that guides an adjustable air
current, generated by a fan and homogenized by means of a polyurethane foam layer. The upflow
intensity is adjusted so that all particles levitate at the minimum possible height over the grid,
thus avoiding friction. The air flow has been produced by a SODECA HCT-71-6T-0.7 fan, and its
uniformity was verified with an anemometer. Local deviations of air flow caudal are found to be
within $\pm 5~\%$. Mean air flows ranged from $2.2$ to $3.2~\mathrm{m/s}$. Particle movement is
limited to a circular region with diameter $L=725 \pm 1~\mathrm{mm}$, located at the center of the
metallic grid and delimited by walls of height $40~\mathrm{mm}$. We used a variable number of disks
ranging from $3$ to $70$ (packing fractions $\phi=0.03$ to $\phi=0.70$).

We have recorded a total of 120 experiments, varying the parameters (density and air current) and
filmed with a Phantom VEO 410L high-speed camera. Each experiment movie contains a series of digital
images at a rate of 250 fps, during $100~\mathrm{s}$. Our images achieve a spatial resolution of
$\sim0.05~\%$ of the particle diameter. From each series of images (movie) we compute 
trajectories, vorticity and other whole-system properties. Trajectories are obtained by means of a
modified version of the algorithm by Crocker and Grier \cite{CG96}.  The typical error for particle
location is $\delta r\lesssim 0.1~\mathrm{pixel}$, (which is near the inherent maximum accuracy in
terms of pixels that can be obtained from the particle tracking method we used
\cite{CG96}). Meanwhile, spinner angular velocities were found using a custom method based on
tracking blade luminosity profiles, A similar experimental configuration has been successfully
employed recently in a series of experiments for studies on active particle dynamics
\cite{Farhadi2018,Workamp2018}.

\section*{Data availability}
The data tables (particle positions and angles during experiments) that support the plots within this paper and other findings of this study
are at the disposal of the reader and available in the following repository: \texttt{https://doi.org/10.5281/zenodo.6121454}.

\section*{Code availability}
The particle tracking codes that were used for this work
are available in the following web link:\\
 \texttt{https://github.com/fvegar/blades}.

\section*{Competing interests}
The authors declare no competing interests.

\section*{Acknowledgments}
We acknowledge funding from the Government of Spain through Agencia Estatal de Investigaci\'on (AEI)
project No. PID2020-116567GB-C22). A.R.-R. also acknowledges financial support from 
Consejer\'ia de Transformaci\'on Econ\'omica, Industria, Conocimiento y Universidades de la Junta de Andaluc\'ia/FEDER 
for funding through project P20-00816 and FSE through post-doctoral grant no. DC00316 (PAIDI 2020). 
F. V. R. is supported by the Junta de Extremadura grant No. GR21091, partially funded by the
ERDF. The authors are indebted to Prof. \'Angel Garcimart\'in, for his important technical and
scientific assistance on the setting up of the laboratory where the experiments of this work were
performed.

\section*{Author contributions statement}
M. L.-C. designed the particles, performed all experiments, wrote most parts of the particle
tracking code and prepared the first version of the manuscript. F. V. R. designed the air table set-up,
conceived the experiments, assisted on the implementation of the particle tracking codes, wrote most
parts of the post-particle-tracking codes for measurement and characterization of the particle
dynamics. F. V. R. and A. R.-R. discussed the results, wrote the paper and co-directed the project.

\section*{Additional information}
\textbf{Publisher’s note} Springer Nature remains neutral with regard to jurisdictional claims in published maps and institutional affiliations.

\bibliography{chiral_diff_regimes}

\end{document}